\documentclass{ws-procs9x6}
\newcommand{\nn}{\nonumber}
\newcommand{\be}{\begin{equation}}
\newcommand{\ee}{\end{equation}}
\newcommand{\ci}[1]{\cite{#1}}
\def\vk{{\bf k}_{\perp}}

\def\vb0{{\bf b}_0}

\def\als{\alpha_s}

\def\gev{\,{\rm GeV}}

\def\xb{\overline{x}}

\def\veps{\varepsilon}
\begin{document}

\title{Spin density matrix elements in vector meson leptoproduction
\footnote{\uppercase{T}his work is supported  in part by the
\uppercase{R}ussian \uppercase{F}oundation for \uppercase{B}asic
\uppercase{R}esearch, \uppercase{G}rant 03-02-16816  and by the
\uppercase{H}eisenberg-\uppercase{L}andau program}}

\author{S.V.\ Goloskokov}

\address{BLTP, Joint Institute for Nuclear Research \\
Dubna 141980, Moscow region, Russia\\
E-mail: goloskkv@thsun1.jinr.ru}

\author{P.Kroll}

\address{Fachbereich Physik, Universit\"at Wuppertal, \\
D-42097 Wuppertal, Germany\\
E-mail: kroll@physik.uni-wuppertal.de}

\maketitle

\abstracts{It is reported on an analysis of vector meson
leptoproduction at small Bjorken-$x$  and large photon virtuality
within an approach that bases on generalized parton distributions
(GPDs). This approach leads to results for cross sections and
spin density matrix elements for $\rho$ and $\phi$ leptoproduction
in fair agreement with experiment.}

Vector meson leptoproduction at small Bjorken-$x$ and large photon
virtuality $Q^2$ factorizes~\cite{Ji97} into a hard
partonic subprocess, vector meson leptoproduction off gluons, and GPDs
describing the collinear emission and reabsorption of gluons from the
proton. The process of interest is dominated by transitions from
longitudinally polarized photons to longitudinally polarized vector
mesons, $\gamma _{L}^{\ast}\to V_{L}$, other transitions are
suppressed by inverse powers of $Q^2$. For experimentally accessible
values of $Q^2$, however, the $L\to L$ transition amplitude, calculated
to leading-twist accuracy, leads to a cross section that exceeds
experiment typically by order of magnitude~\cite{mpw}. Moreover, for
$Q^2$ of the order of, say, $10\,\gev^2$ other transitions can not be
ignored as data on spin density matrix elements reveal. A calculation
of the transition amplitudes involving transversely polarized photons
and/or vector mesons ($T$) in collinear factorization is
problematic since infrared singularities occur indicating a break down
of factorization~\cite{mp}.

We are reporting here on work in progress in which the GPD approach is
modified by employing the modified perturbative approach
\cite{sterman} in the calculation of the hard subprocess: the
transverse degrees of freedom of the quarks building up the vector
meson are retained and Sudakov suppressions are taken into account.
This method provides a regularization scheme for the infrared
singularities mentioned above. In passing we note that our approach
bears resemblance to models proposed in \cite{martin} where, up to
occasional corrections, the GPDs are replaced by the usual gluon
distributions.

The gluon contribution to the leptoproduction amplitude, calculated at
the momentum transfer $t\simeq t_{\rm min}\simeq 0$ for a proton with
positive helicity, reads
\begin{eqnarray}
{ M}_{\mu'+,\mu +} &=& \frac{e}{2}\, { C}_V\,
         \int_0^1 \frac{d\xb}
                   {(\xb+\xi)(\xb-\xi + i{\veps})}\nn\\
        &\times& \left[\,{A}^V_{\mu'+,\mu +}\,
         + { A}^V_{\mu'-,\mu -}\,\right]\,
                                   H^g(\xb,\xi,t\simeq 0)\,,
\label{amptt-nf-ji}
\end{eqnarray}
where $\mu$ ($\mu'$) denotes the helicity of the photon
(meson). Proton helicity flip and contributions from the GPD
$\widetilde{H}^q$ which is much smaller than the GPD $H^g$ at small
$\xb$, are neglected in (\ref{amptt-nf-ji}). $\xb$ is the average
momentum fraction of the gluons and the skewness $\xi$ is related to
Bjorken-$x$ by $\xi\simeq x_{Bj}/2$. The flavor factors are
$C_{\rho}=1/\sqrt{2}$ and ${ C}_{\phi}=-1/3$.

The hard scattering amplitudes for the various helicity
configurations are (the signs now label gluon helicities)
\be
A^V_{\mu'+,\mu +}\, = \,\frac{2\pi \als(\mu_R)}
           {N_c} \int_0^1 d\tau\int \frac{d k^2_\perp}{16\pi^2}
            \Psi_V(\tau,k^2_\perp)\;
            K_{\mu'\mu}(\tau,\xb,\xi,k^2_\perp,Q^2)\,.
\label{tt-ji}
\ee
The hard scattering kernels $K$ are calculated from the relevant
Feynman graphs at $t\simeq 0$ (see for instance \cite{HW}), only the
factors of $\sqrt{-t}$ required by angular momentum conservation, are
kept. The amplitude (\ref{tt-ji}) is Fourier transformed to the impact
parameter space where the variable $\bf b$ is canonically conjugated
to ${\bf k_\perp}$. Multiplication with the $b$-space version of the
Sudakov factor~\ci{sterman} completes the specification of the
subprocess amplitude.

In the calculation of the subprocess a spin wave function for the
vector meson is used which takes into account the quark transverse
momentum, $\vk$, linearly, and, hence, one unit of orbital angular
momentum in a covariant manner~\cite{koerner}. For the momentum-space
meson wavefunction $\Psi_V$ a simple Gaussian form is adopted~\cite{jak93}
 \be
   \Psi_V(\vk,\tau)\,=\, 8\pi^2\sqrt{2N_c}\, f_V a^2_V
       \, \exp{\left[-a^2_V\, \frac{\vk^{\,2}}{\tau\bar{\tau}}\right]}\,,
\label{wave-l}
\ee
where $\tau$ is the fraction of the meson momentum the quark carries
($\bar{\tau}=1-\tau$). In the numerical evaluation of meson
leptoproduction the following values for the decay constants and the
transverse size parameters are used: $f_\rho = 0.216\,\gev$,
$f_\phi = 0.237\,\gev$, $ a_\rho= 0.52\, \gev^{-1}$ and
$a_\phi= 0.45\, \gev^{-1}$.

The GPD is modelled from a double distribution~\cite{mus99} for which
the following ansatz is exploited:
\be \
f^g(\beta,\alpha,t\simeq 0)) \sim
\frac{\big[(1-|\beta|)^2-\alpha^2\big]^n}{(1-|\beta|)^{2n+1}}\,
g(\beta)\,.
\ee
An appropriate integral over $f^g$ leads to the GPD~\ci{mus99}.
For the gluon distribution $g(\beta)$ we take the CTEQ5M
parameterization~\cite{CTEQ}. The cases $n=1$ and 2 lead to nearly the
same results.

\begin{figure}[b]
\begin{center}
\includegraphics[width=4.5cm,bbllx=37pt,bblly=351pt,bburx=529pt,
bbury=742pt,clip=true]{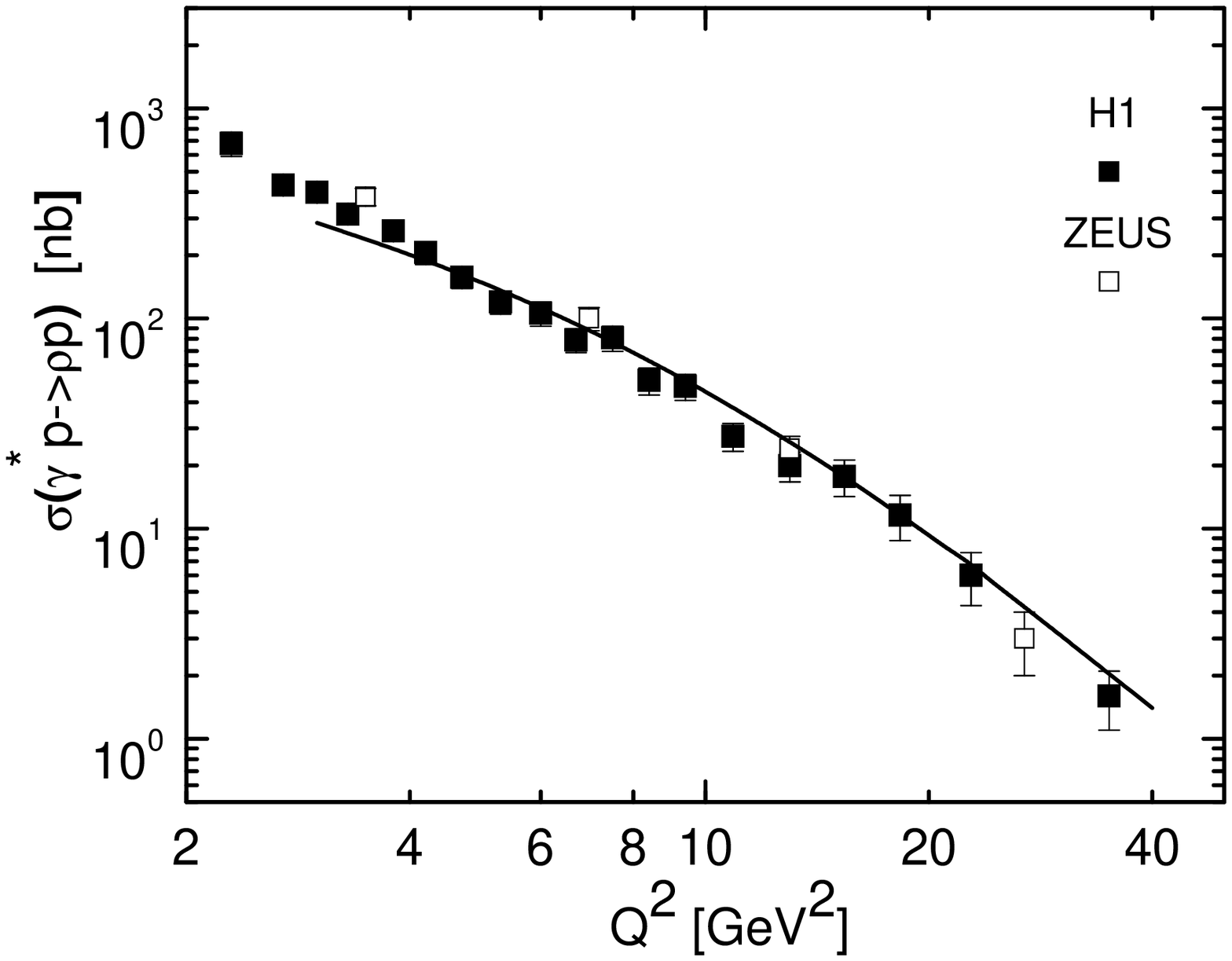} \hspace*{1em}
\includegraphics[width=4.5cm,bbllx=30pt,bblly=333pt,bburx=545pt,
bbury=742pt,clip=true]{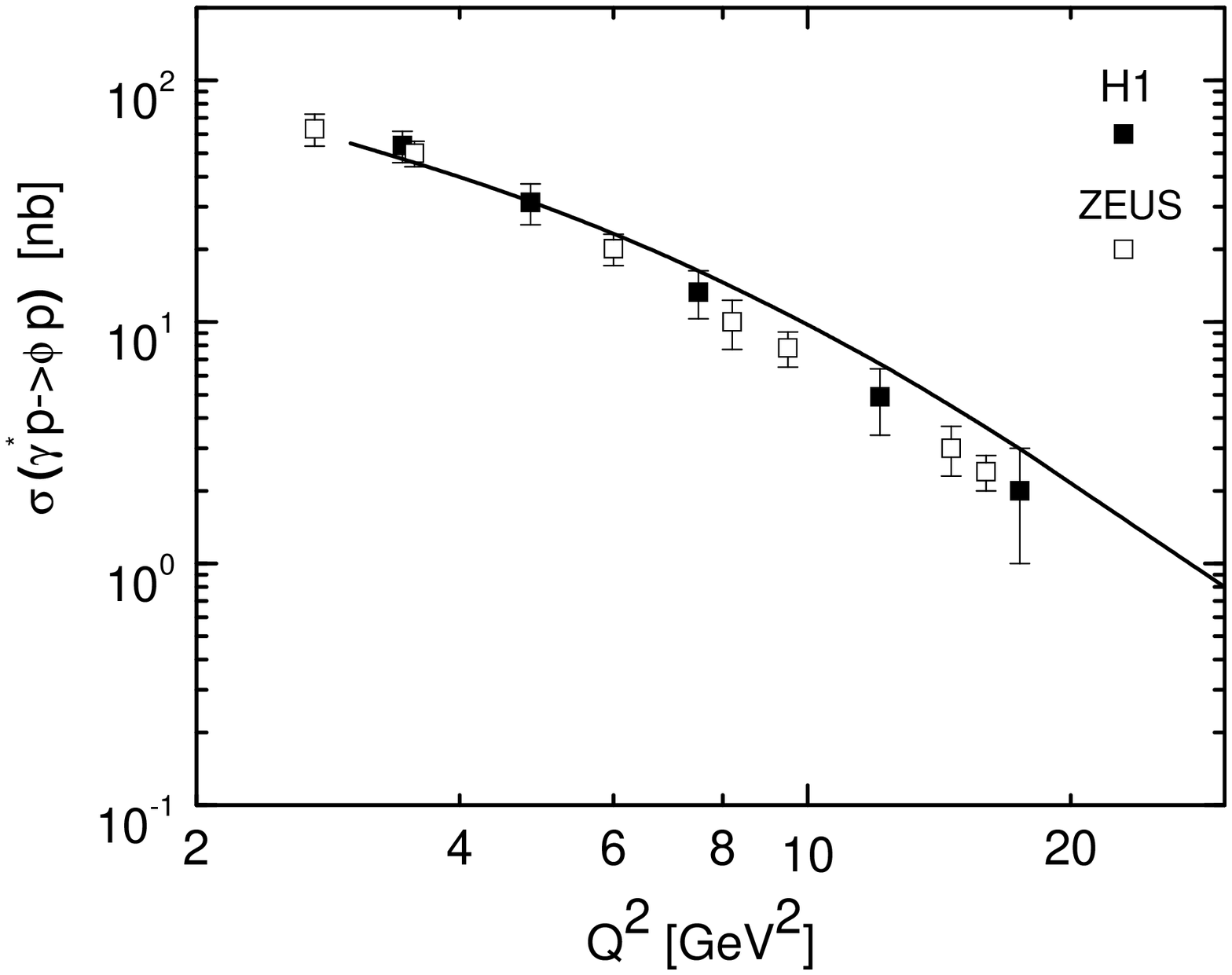} \caption{The integrated cross
section for $\gamma^*p\to \rho p$
  (left) and $\gamma^* p\to \phi p$ (right)  at
  $W\simeq 75\, \gev$. Data taken from \protect\ci{h1} (filled squares) and
  \protect\ci{z95}(open squares). The solid lines represent our results.}
\label{fig:cross}
\end{center}
\end{figure}

We have evaluated the amplitudes for $L\to L$, $T\to L$ and $T\to T$
transitions at small $t$ and neglected the $L\to T$ and $T\to -T$ ones
since they are strongly suppressed in our approach. The three
amplitudes are multiplied by exponentials $\exp[-B_it/2]$ in order to
make contact with experiment (or to extrapolate to $t=0$).
The  integrated cross  sections for $\gamma^* p \to V p$
are shown in Fig.\ \ref{fig:cross}. Good agreement with experiment is to be observed
for $\rho$ and $\phi$ mesons.

\begin{figure}[h]
\begin{center}
\includegraphics[width=4.4cm,bbllx=79pt,bblly=300pt,bburx=545pt,
bbury=739pt,clip=true]{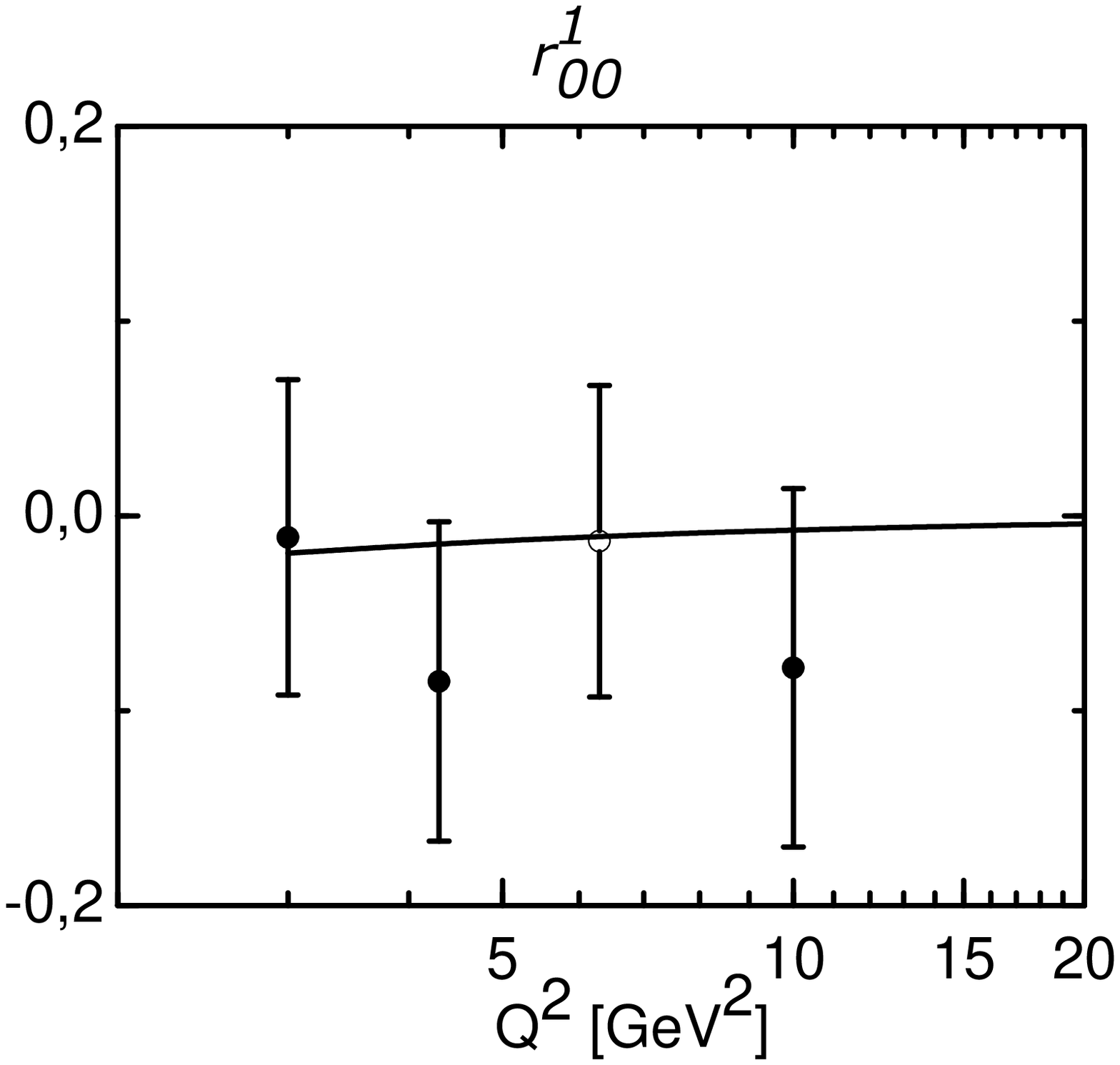} \hspace*{1em}
\includegraphics[width=4.5cm,bbllx=75pt,bblly=303pt,bburx=555pt,
bbury=741pt,clip=true]{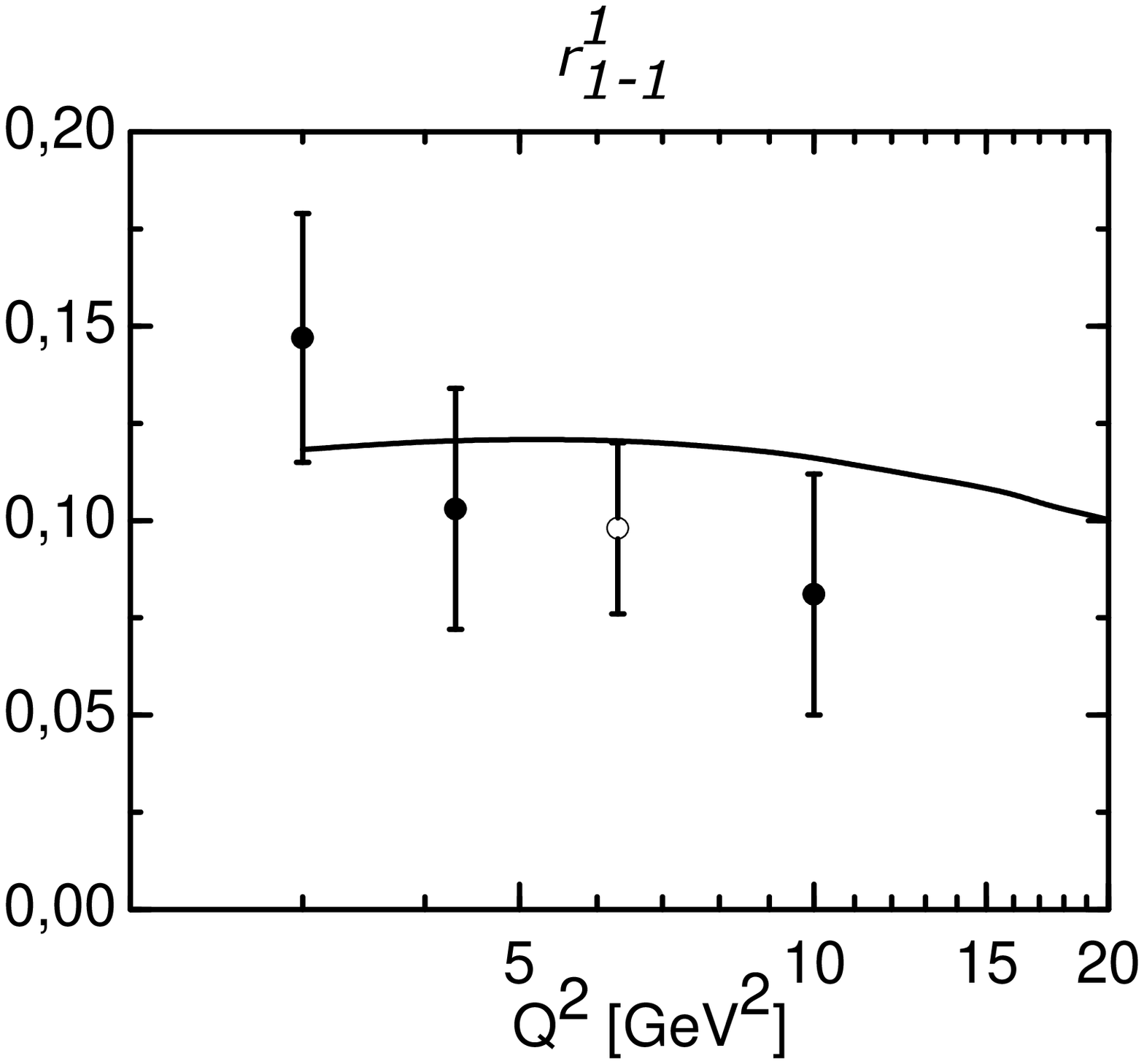} \caption{Sample spin density
matrix elements of electroproduced $\rho$ at $W=75\,\gev$ and
$\langle -t\rangle=0.15\,\gev^2$. Data from \protect\ci{h1}
(filled circles) and \protect\ci{zeus99} (open circles), are
compared to our results (solid lines). } \label{fig:sdme}
\end{center}
\end{figure}

In Fig.\ \ref{fig:sdme} two examples for predictions of spin
density matrix elements are shown. Our results reproduce quite well the
$Q^2$ dependence of the spin density matrix obtained at HERA
experiments \cite{h1,zeus99}.

In summary, the GPD approach combined with the modified perturbative
approach applied to the partonic subprocess allows for a calculation
of various cross section and spin density matrix elements at large $Q^2$
and small Bjorken-$x$.  Fair agreement between theory and experiment
for light vector meson leptoproduction is found.
It is important to extend our results to lower
energies but larger Bjorken-$x$. This is the HERMES and COMPASS region
where the quark GPDs have be taken into account.
Preliminary data for the spin density matrix
elements have been presented by COMPASS
at this conference \cite{compass}.

\end{document}